# Effect of time and storing conditions on iron forms in ferrous gluconate and Ascofer®

R. Gozdyra, S. M. Dubiel[*] and J. Cieślak

Faculty of Physics and Computer Science, AGH university of Science and Technology, al. A. Mickiewicza 30, PL-30-059 Krakow, Poland

## Abstract

Antianemic medicament Ascofer® and ferrous gluconate, its basic iron bearing ingredient, were studied with the use of Mössbauer spectroscopy. Room temperature spectra gave a clear evidence that two phases of iron were present viz. ferrous ($Fe^{2+}$) as a major one with a contribution of ~85 ±5 %, and ferric ($Fe^{3+}$) whose contribution was found to be ~15±5 %. However, the actual values of the contributions of the two kind of the iron ions in Ascofer® depend on sample's age: the abundance of $Fe^{2+}$ ions increases with time by ~10% after 51 months, while that of $Fe^{3+}$ decreases by the same amount. This means that an internal reduction of $Fe^{3+}$ ions takes place. Ferrous ions were shown to occupy at least two different sites. In Ascofer®, the relative abundance of the two sites does not depend on the age of sample, while in the gluconate the population of site 1 increases and that of site 2 decreases with the age of the sample.

Key words: Ferrous gluconate, ferrous and ferric ions, Mossbauer spectroscopy

[*]Corresponding author: dubiel@novell.ftj.agh.edu.pl

## 1. Introduction

Iron deficiency anemia is the most common form of anemia. In case of mild and medium stage of the disease, its treatment involves taking iron supplements. To be absorbed, dietery iron must be in its ferrous ($Fe^{2+}$) form, which is soluable. Consequently, the iron supplements for oral inake are usually in form of ferrous salts like: ferrous fumarate e. g. Feostat®, Ferret® or Femiron®; ferrous sulfate e. g. Feosol®, Feratab® or FerSul®, and ferrous gluconate e. g. Fergon®, Simron® or Ferralet®. Like all other medicaments also those for the treatment of anemia, have their characteristic expiration date and they should be stored in appropriate conditions like a dark and dry place. The purpose of this study was to see whether or not the storing conditions affect the form of iron and what happens with iron when the expiration date has been overrun. As our research target we have chosen Ascofer®, a medicament produced by a Polish farmaceutical firm, Espefa®, based in Kraków. It is manufactured in form of dragee having a core and a shell. The core contains 200 mg ferrous gluconate (23.2 mg elemental Fe) mixed with starch, talc, ascorbic and stearin acids, and propylene glycol. The shell is composed of talc, saccharine, gelitin, magnesium oxide, arabic gum and cochineal red. Among these constituents only the ferrous gluconate shold contain iron. According to the producer, the medicament should be stored in a dry and dark place at temperature between 15 °C and 25 °C, and not taken after the expiration date which occurs 2 years after the fabrication date.

The ferrous gluconate itself was studied with the Mössbauer Spectroscopy (MS) in the past [1-3]. Authors of Refs. 1 and 2 analyzed measured spectra with one doublet having spectral parameters charactersitic of $Fe^{2+}$ ions in the high-spin state, while authors of Ref. 3 interpreted their spectra in terms of two doublets ascribed to $Fe^{2+}$ ions occupying two different sites. They also found that both kind of the gluconate they investigated viz. a commercial one and a freshly prepared in a chemical laboratory were contaminated with ~10 at% of iron in form of $Fe^{3+}$. In view of the above, one of our aims was to verify which of the two approaches applied in the spectra analysis is more appropriate. This is an important issue in the light of still unknown crystal structure of the iron gluconate. It should be added that Mössbauer spectroscopy has been recently successfully applied in investigations of various industrial samples of multiple vitamins and dietary supplements with iron in different ferrous forms like fumarate, bisglycinate chelate or ferrous sulfate [4-7].

## 2. Experimental

### 2.1. Samples

$^{57}Fe$ Mössbauer spectra were recorded both on particular components of Ascofer® i.e. (a) ferrous gluconate, (b) talc, (c) saccharine, (d) gelatin, (e) magnesium oxide, (f) arabic gum and (d) cochineal red, as well as on Ascofer® itself.

Among the components only the ferrous gluconate was expected to contain iron. However, the iron was also found to be present in the talc. Its room temperature spectrum could have been well fitted with one singlet and one doublet. Their spectral parameters are displayed in Table 1.

**Table 1**
The best-fit spectral parameters of a room temperature spectrum recorded on the talc (*a* – abundance; *IS* – isomer shift; *Γ* – line width at half maximum; *QS* – quadrupole splitting)

|        | singlet | doublet |
|--------|---------|---------|
| A [%]  | 19.2    | 80.8    |

| IS [mm/s] | 0.32 | 1.12 |
|---|---|---|
| Γ [mm/s] | 0.34 | 0.34 |
| QS [mm/s] | - | 2.72 |

The values of the spectral parameters of the doublet are close to those of FeSO$_4$ [6]. They are also similar to those characteristic of Fe$^{2+}$ ions at site 2 – see Table 2 below. The latter feature may introduce some uncertainty as far as the relative contribution of Fe$^{2+}$ ions at site 2 is concerned, as the spectral parameters of the latter have similar values.
Concerning the ferrous gluconate, two samples A and B were investigated. They were prepared by various manufacturers.

## 2. 2. Mössbauer spectra and their analysis

$^{57}$Fe-site Mössbauer spectra were recorded at room temperature in a transmission geometry using a standard spectrometer with a drive working in a sinusoidal mode. Gamma rays of 14.4 keV energy were supplied by a $^{57}$Co/Rh source. All investigated samples were in form of powder that in the case of Ascofer was produced by attrition a dragee in an agate mortar. Investigated samples of the ferrous gluconate had a mass of 200 mg and the case of Ascofer® one dragee was used as a sample.
All spectra of the ferrous gluconate and of Ascofer® had similar shape i.e. asymmetric doublet with a small bumb close to the inner part of the left-hand line – see Fig. 2. The shape reflects the two-phase composition known from the previous studies [1-3]. The major ferrous phase is represented by the two intensive lines from which one is centered at ~-0.5 mm/s and the other one at ~2.5 mm/s. The minor ferric phase is also represented by two lines (doublet) from which the left-hand one is hidden under the much more intensive left-hand line associated with the ferrous phase, while the other one can be seen in form of the bumb centered at ~0.5 mm/s. In the previous studies the subspectrum due to the ferrous phase was treated either as a doublet [1] or two doublets [3]. As one of the aims of this study was to verify whether or not Fe$^{2+}$ ions ocupy two different sites as claimed in [3], the subspectrum was fitted with the quadroupole distribution method i.e. *a priori* no assumption on the number of sites occupied by Fe$^{2+}$ ions was made. The subspectrum representing the ferric phase was treated as one doublet. The asymmetry of the major quasi-doublet, which according to [3] does not originate, as suggested in [1], from the Goldanskij-Karyagin effect, was here accounted for by assuming a linear correlation between the isomer shift, *IS*, and the quadroupole splitting, *QS*. Using such procedure all the measured spectra could have been successfully fitted. All values of *IS* are given relative to metallic iron.

## 3. Results and discussion

### 3.1. Ferrous Gluconate

#### 3.1.1. Effect of origin

The spectra measured on samples A and B together with the corresponding distributions of *QS* can be seen in Figs. 2 and 3, respectively. It is obvious that the distribution of *QS* in both samples is not symmetric which means that Fe$^{2+}$ ions occupy more than one site. Assuming the Gaussian distribution of *QS* for each site, two Gaussians must be taken into account in order to correctly fit the overall distribution curve. The centre of each Gaussian gives the value of *QS* asociated with the particular site and its relative area the relative population of this site. Spectral parameters obtained with such procedure are presented in Table 2. As follows from this table, in both investigated samples i. e. A and B iron exists in two forms viz. as a ferrous ($Fe^{2+}$) and as a ferric ($Fe^{3+}$). The former with the abundance of ~85 – 88% is the major fraction while the latter is a minor fraction and is basically an undesired phase. The data displayed in Table 2 also give a clear evidence that $Fe^{2+}$ ions occupy, at least, two different sites that differ in values of the quadrupole splitting, *QS*, and abundances. As the origin of the difference is not known, we will assume, following [3] that the two variations of $Fe^{2+}$ ions occupy sites with slightly different symmetry because they have different *QS*-values. The data presented in Table 2 give evidence that the actual phase composition of the investigated samples A and B is meaningfully different while the values of the spectral parameters ascribed to $Fe^{2+}$ and $Fe^{3+}$ ions are rather similar in both samples. That means that the difference in the relative phase composition is of technological origin. Such conclusion can be further supported by the results discussed in [3], according to which the contribution of $Fe^{3+}$ ions in the samples studied by those authors was ~10 at%. Also the ratio between the abundances of the $Fe^{2+}$ ions present at the two sites seems to be strongly dependent on the technology of sample production: that determined in this study is equal to ~3 which is much less than ~5 as found elsewhere [3].

**Table 2**
The best-fit spectral parameters as obtained by fitting room temperature spectra recorded on ferrous gluconate samples A and B. Abundances, *<a>* and *a* are in [%], isomer shifts, *IS*, quadruople splitings, *QS*, distribution width, *G* and line width at half maximum, *Γ* in [mm/s].

|  | $Fe^{2+}$ | | | | | | | | | $Fe^{3+}$ | | | |
|---|---|---|---|---|---|---|---|---|---|---|---|---|---|
|  | average | | | Site 1 | | | Site 2 | | | | | | |
| sample | <a> | <IS> | <QS> | a | QS | G | a | QS | G | a | IS | QS | Γ |
| A | 87.9 | 1.21 | 2.93 | 74.6 | 3.04 | 0.28 | 25.4 | 2.68 | 0.42 | 12.1 | 0.38 | 0.89 | 0.63 |
| B | 84.7 | 1.21 | 2.90 | 72.8 | 3.04 | 0.32 | 27.2 | 2.52 | 0.52 | 15.3 | 0.41 | 0.80 | 0.80 |

### 3.1.1. Effect of time (aging)

To study the effect of aging, three Mössbauer spectra were recorded at 295 K on a sample of gluconate kept in a dark and dry place. Spectrum 2 was recorded seven months after spectrum 1, and spectrum 3 fourteen months later than spectrum 2. In other words there is a time difference of twenty one months between spectrum 1 and 3. Spectral parameters obtained from these spectra are presented in Table 3 and in Fig. 6. As can be seen the average abundance of $Fe^{2+}$ fraction does not show any systematic behaviour while the relative abundance at particular sites does. Concerning the latter, the population of $Fe^{2+}$ ions on site 1 has incread with time from ~73 % to ~92 %, while that on site 2 has decreased with time from ~27 % to ~8 % after 21 months. In addition, ~4% of a new fraction (site 3) has appeared after 21 months of aging.

**Table 3**
The best-fit spectral parameters relevant to the aging process as obtained by fitting room temperature spectra recorded on ferrous gluconate samples with various age. Age is in months, relative abundances, *<a>*, *a* and $a_{1,2,3}$ are in [%],

|  | $Fe^{3+}$ | $Fe^{2+}$ | | | |
|---|---|---|---|---|---|
| age | a | <a> | $a_1$ | $a_2$ | $a_3$ |
| 0 | 15.3 | 84.7 | 72.8 | 27.2 | - |
| 7 | 12.1 | 87.9 | 74.6 | 25.4 | - |

| 21 | 17.8 | 89.4 | 91.7 | 8.3 | 4.1 |

## 3. 2. Ascofer®

The $^{57}$Fe Mössbauer spectrum of Ascofer® should, in principle, be identical with that of the ferrous gluconate recorded in the same conditions. A possible difference between the spectra recorded on different samples of Ascofer® may originate in a use of (a) ferrous gluconate and/or (b) talc of different origins for its fabrication. As it has been described above, the samples A and B of the ferrous gluconate differ by ~5% in the relative abundance of $Fe^{2+}$ ions, hence a difference of this order in the relative amount of these ions can be expected in Ascofer®. To verify whether or not this is true, we have recorded room temperature spectra on four different samples produced in 2007. The spectral parameters obtained from the spectra are presented in Table 4. It can readily be seen that, in fact, the abundances of $Fe^{2+}$ ions, $<a>$, fall into two groups: (I) with $<a>$ between ~82-83%, hence close to that found for the ferrous gluconate B, and (II) with $<a>$ between ~88-91%, hence close to that revealed for the ferrous gluconate A. A lack of an exact match may be due to the presence of additional $Fe^{2+}$ ions in Ascofer due to the talc which, in general, may have a different origin, hence different amount of iron.

**Table 4**
The best-fit spectral parameters as obtained by fitting room temperature spectra recorded on four samples of Ascofer® all produced in the same year but having different series. Abundances, $<a>$ and $a$ are in [%], isomer shifts, *IS*, quadruople splittings, *QS*, and line width at half maximum, $\Gamma$ in [mm/s].

| | $Fe^{2+}$ | | | | | | | | | | | $Fe^{3+}$ | | | |
|---|---|---|---|---|---|---|---|---|---|---|---|---|---|---|---|
| | average | | | Site 1 | | | Site 2 | | | Site 3 | | | | | | |
| sample | $<a>$ | $<IS>$ | $<QS>$ | a | IS | QS | a | IS | QS | a | IS | QS | a | IS | QS | $\Gamma$ |
| 1 | 91.2 | 1.22 | 2.90 | 81.6 | 1.09 | 3.04 | 13.7 | 1.06 | 2.54 | 4.7 | 1.02 | 1.92 | 8.8 | 0.28 | 1.04 | 0.51 |
| 2 | 87.8 | 1.22 | 2.99 | 93.8 | | 3.04 | 6.2 | | 2.42 | - | - | - | 12.2 | 0.33 | 0.92 | 0.64 |
| 3 | 82.6 | 1.22 | 2.94 | 83.6 | | 3.04 | 13.6 | | 2.48 | 2.8 | | 1.84 | 17.4 | 0.27 | 0.80 | 0.82 |
| 4 | 81.8 | 1.22 | 2.96 | 89.9 | | 3.04 | 5.8 | | 2.46 | 4.3 | | 1.98 | 18.2 | 0.27 | 0.77 | 0.59 |

### 3. 2. 1. Effect of time (aging)

To see whether or not the age of Ascofer® has any effect on the iron phases, two different series of the medicament were first considered: (i) 150503 and (ii) 150505. Content of $Fe^{2+}$ ions gives hint that the series 150503 might have been produced with the ferrous gluconate B, while the series 150505 with the ferrous gluconate A. The Mössbauer spectra were recorded on the series 150503 27, 30 and 51 months after its fabrication, while those on the series 150505 4 and 25 months after its fabrication. The abundances of $Fe^{2+}$ and $Fe^{3+}$ ions as determined from the distributions of the quadrupole splittings are shown in Figs. 6 and 7.
The figures give a clear evidence on the aging effect. The abundance of $Fe^{2+}$ ions increases with time, while that of $Fe^{3+}$ decreases. In particular, in the series 150503 the amount of the $Fe^{2+}$ ions has increased by ~11% during 24 months at the cost of $Fe^{3+}$ ions. This effect can be understood as due to a reduction process i.e. $Fe^{3+}$ ions change their valence into $Fe^{2+}$. This also means that $Fe^{3+}$ ions are an integral part of the ferrous gluconate structure. On the other hand, as shown in Fig. 8, the distribution of $Fe^{2+}$ ions over the site 1 and site 2 is rather insensitive to the aging. This is in contrast to the corresponding behaviour found in the pure ferrous gluconian where, as illustrated in Fig. 5, a systematic redistribution of $Fe^{2+}$ ions over the site 1 and site 2 has taken place in the course of time. The difference may follow from the fact that in Ascofer the ferrous gluconate exists as a core surrounded by a shell, hence it is protected from the influence of external factors such as light or temperature.

To further study the effect of aging, room temperature Mössbauer spectra were recorded on several samples of Ascofer® belonging to different series. Their age spans over a period of 51 months. The spectra were analyzed as before, and the spectral parameters relevant to the aging are shown in Figs. 9 and 10.

A positive correlation between the population of $Fe^{2+}$ ions, $<a>$, and the sample's age can be seen. It seems there are two branches in Fig. 9 – one presented with full and the other with open symbols. The former could represent Ascofer® fabricated with the ferrous gluconate A and the latter with B. This correlation is accompanied by a negative correlation between the abundance of $Fe^{3+}$ ions and aging time – see Fig. 10. From the two correlations follows that the increase of the population of $Fe^{2+}$ ions occurs at the cost of that of $Fe^{3+}$ ones. In other words, the latter ions are reduced into the former.

### 3. 3. 3 Effect of sunshine and air

To see whether the sunshine has any measurable effect on the iron forms in Ascofer®, two series one freshly fabricated (2007) and another one 4 years old (2003) kept in an original blister were exposed during 10 months to sunshine. In addition, one sample from a freshly fabricated Ascofer® series 20105 was powdered and kept in contact with air during the same period of time. The values of the spectral parameters obtained from these spectra are displayed in Table 6.

**Table 6**
The best-fit spectral parameters of Ascofer® samples produced in 2003 and 2007 kept in original blisters before and after (a) 10 months exposure to sunshine as well as those obtained for Ascofer® series 20105 exposed to air in form of powder (20105p).

| | $Fe^{2+}$ | | | | | | | | | | | | $Fe^{3+}$ | | | |
|---|---|---|---|---|---|---|---|---|---|---|---|---|---|---|---|---|
| | average | | | Site 1 | | | Site 2 | | | Site 3 | | | | | | |
| sample | $<a>$ | $<IS>$ | $<QS>$ | a | IS | QS | a | IS | QS | a | IS | QS | a | IS | QS | Γ |
| 2003 | 89.5 | 1.215 | 2.90 | 91.7 | 1.09 | 3.06 | 5.6 | 1.06 | 2.22 | 2.7 | 1.02 | 1.86 | 10.5 | 0.265 | 1.06 | 0.77 |
| 2003a | 83.3 | 1.23 | 3.00 | 92.5 | | 3.06 | 7.5 | | 2.26 | - | - | - | 16.7 | 0.27 | 0.82 | 0.72 |
| 2007 | 91.2 | 1.22 | 2.90 | 81.6 | | 3.04 | 13.7 | | 2.54 | 4.7 | | 1.92 | 8.8 | 0.28 | 1.04 | 0.55 |
| 2007a | 81.7 | 1.22 | 2.90 | 83.1 | | 3.04 | 11.1 | | 2.52 | 5.8 | | 2.10 | 18.3 | 0.27 | 0.80 | 0.84 |
| 20105 | 80.3 | 1.22 | 2.94 | 88.0 | | 3.04 | 7.6 | | 2.42 | 4.3 | | 1.96 | 19.7 | 0.27 | 0.80 | 0.68 |
| 20105p | 73.7 | 1.22 | 2.98 | 88.9 | | 3.02 | 7.9 | | 2.44 | 3.1 | | 1.94 | 26.3 | 0.34 | 0.85 | 0.67 |

One can see for the samples produced in 2003 and 2007 that in those exposed to the sunshine the population of $Fe^{2+}$ ions has decreased by ~6% in the older sample and by ~9.5% in the younger sample, while that of $Fe^{3+}$ ions has correspondingly increased in both samples. Such behaviour, rather unexpexted, means that an oxidation process has occurred. Probably the blisters in which the dragees were kept were not air proof. Similar change, this time expected, can be seen in the sample in form of powder exposed to air. As follows from the last two lines of Table 6, about 10% of $Fe^{2+}$ ions have been oxidized into $Fe^{3+}$ ones with practically no change in their distribution over the 3 sites. The latter observation means that all these sites occupied by $Fe^{2+}$ ions are equivalent at least as far as the oxidation process is concerned.

### 4. Summary and conclusions

Using $^{57}$Fe-site Mössbauer spectroscopy, samples of a pure ferrous gluconate and those of Ascofer® medicament were investigated. In the former iron exists as (1) $Fe^{2+}$ ions and (2) $Fe^{3+}$ ions. The former constitutes a major phase with a population between ~83 and ~85 % depending on the origin of the gluconate. The $Fe^{2+}$ ions are distributed at least over two sites with a population of ~73-75% for site 1 and ~17-15% for site 2. However, the actual distribution depends on the age of the sample and it decreases with time for site 1 and increases for site 2. In Ascofer, iron exists also in the same forms as in the ferrous gluconate. However, some small amount of iron was also found in talc which is a constitutiong ingredient of Ascofer®.

The actual amount of different forms of iron in Ascofer® was found to depend on the age of the medicament. The population of $Fe^{2+}$ ions shows an increasing tendency as a function of age while that of $Fe^{3+}$ ions an decreasing one. This means that $Fe^{3+}$ ions, which are rather undesired, undergo a reduction process and transform in the course of time into $Fe^{2+}$ ions. However, based on the present study one cannot give any hint on how does this process occur, but it seems that $Fe^{3+}$ ions are intergral part of the structure of the gluconate. The abundances of the two forms of iron present in Ascofer® can also be affected by sunshine and air. In both cases an oxidation process of $Fe^{2+}$ ions into $Fe^{3+}$ ones happens which is reflected by the increase of the population of $Fe^{3+}$ ions accompanied by the decrease of $Fe^{3+}$ ones. Finally, there is some evidence that $Fe^{2+}$ ions may occupy a third site.

**Acknowledgement**

The study was carried out with a financial support of the Ministry of Science and Higher Education, Warszawa. Espefa® is thanked for a supply of a major part of the samples investigated in this study.

**Figures**

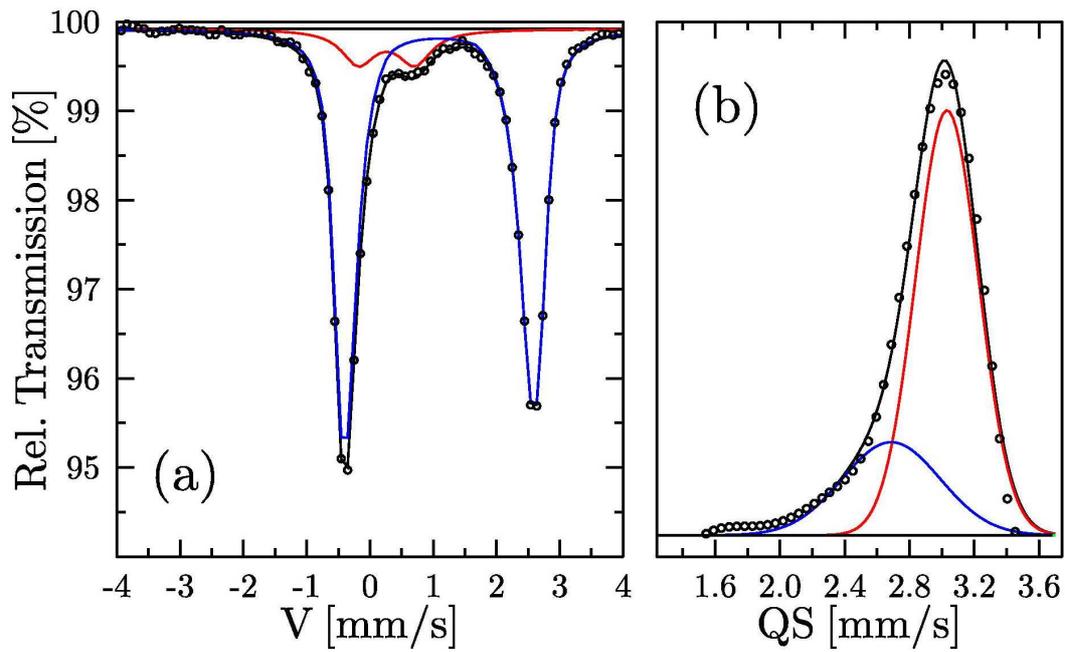

**Fig. 1**
(a) $^{57}$Fe Mössbauer spectrum recorded at 295 K on sample A of the ferrous gluconate, and (b) the corresponding distribution of the quadrupole splitting derived from the subspectrum ascribed to $Fe^{2+}$ ions (major doublet in (a)).

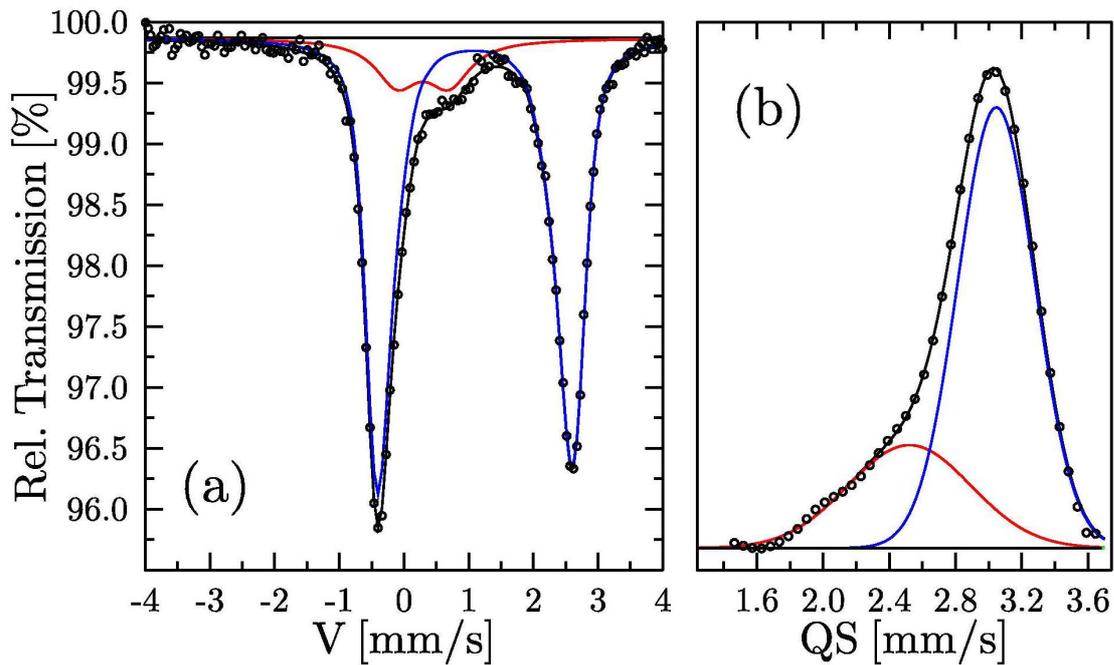

**Fig. 2**
(a) $^{57}$Fe Mössbauer spectrum recorded at 295 K on sample B of the ferrous gluconate, and (b) the corresponding distribution of the quadrupole splitting derived from from the subspectrum ascribed to $Fe^{2+}$ ions (major doublet in (a)).

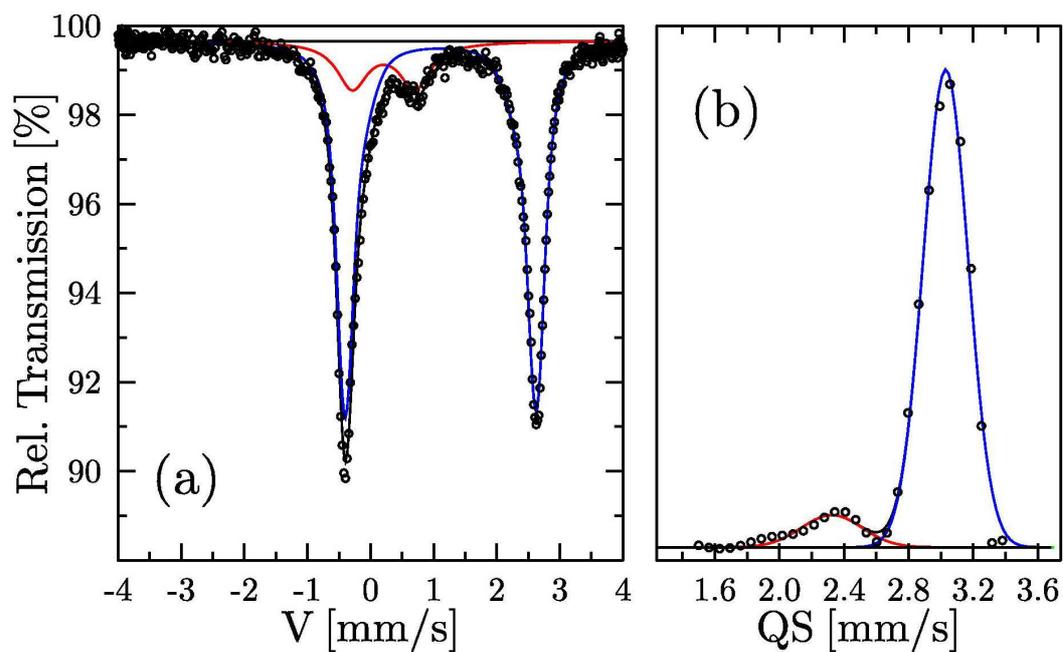

**Fig. 3**
(a) $^{57}$Fe Mössbauer spectrum recorded at 295 K on sample of the ferrous gluconate kept away from the sunshine, and (b) the corresponding distribution of the quadrupole splitting derived from from the subspectrum ascribed to $Fe^{2+}$ ions (major doublet in (a)).
.

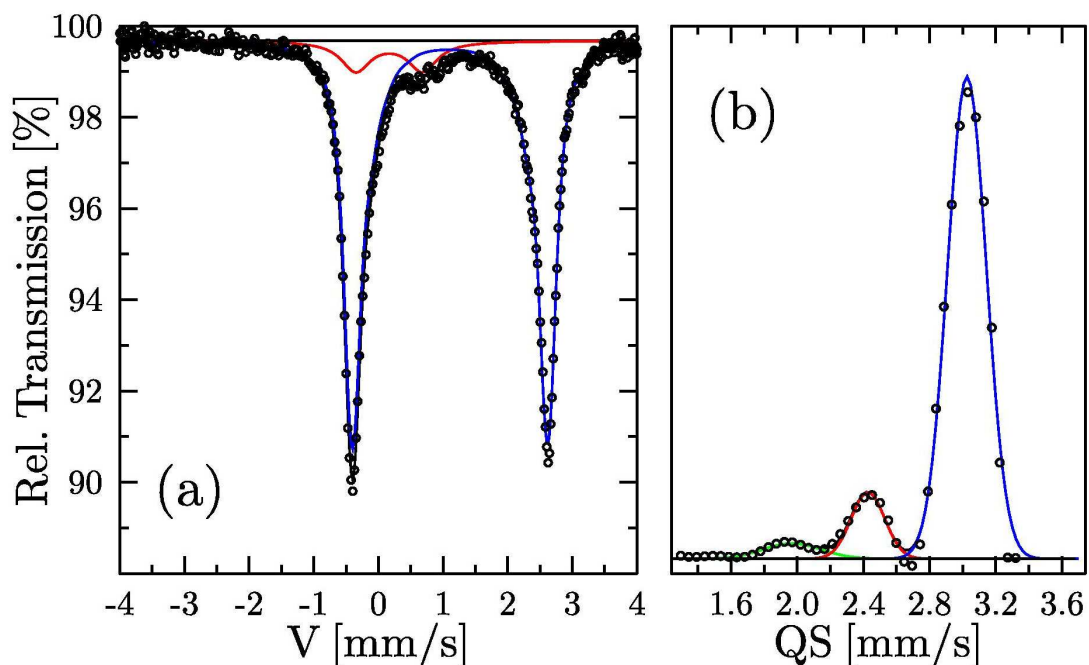

**Fig. 4**
(a) $^{57}$Fe Mössbauer spectrum recorded at 295 K on sample of the ferrous gluconate kept for a year in the sunshine, and (b) the corresponding distribution of the quadrupole splitting derived from from the subspectrum ascribed to $Fe^{2+}$ ions (major doublet in (a)).
.

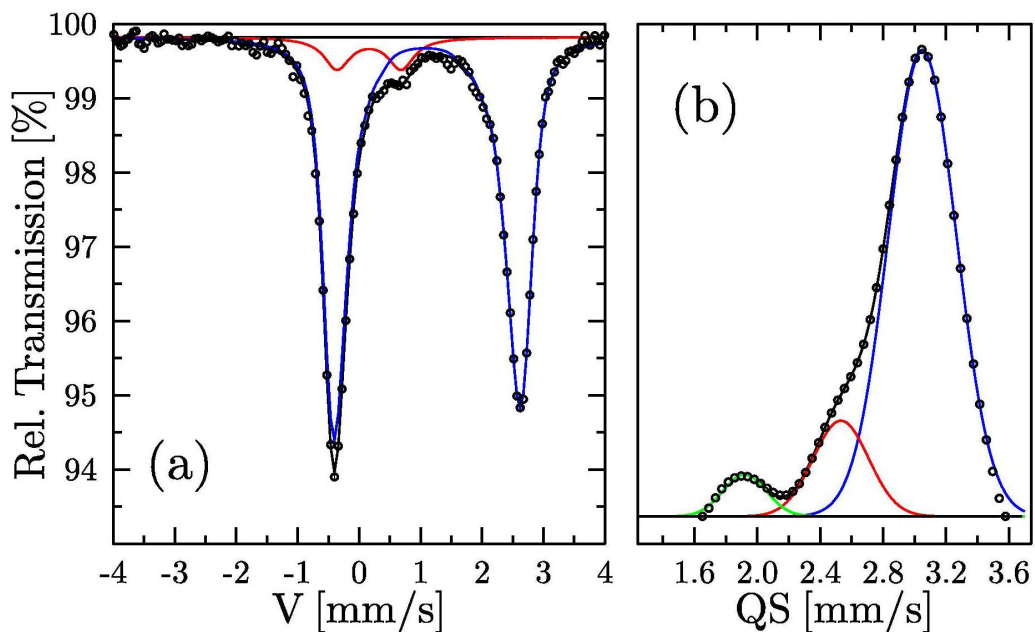

**Fig. 5**
(a) $^{57}$Fe Mössbauer spectrum recorded at 295 K on sample of Ascofer few months old, and (b) the corresponding distribution of the quadrupole splitting derived from the subspectrum ascribed to $Fe^{2+}$ ions (major doublet in (a)). In the latter 3 different sites are indicated.

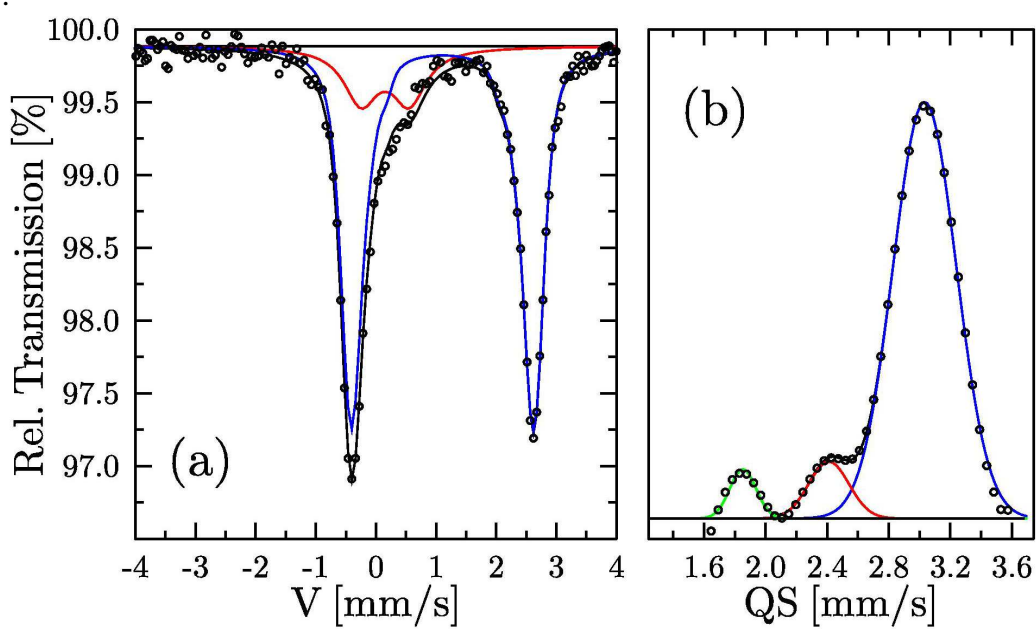

**Fig. 6**
(a) $^{57}$Fe Mössbauer spectrum recorded at 295 K on sample of Ascofer few months old (different series than the one shown in Fig. 5), and (b) the corresponding distribution of the quadrupole splitting derived from the subspectrum ascribed to $Fe^{2+}$ ions (major doublet in (a)). In the latter 3 different sites are indicated.

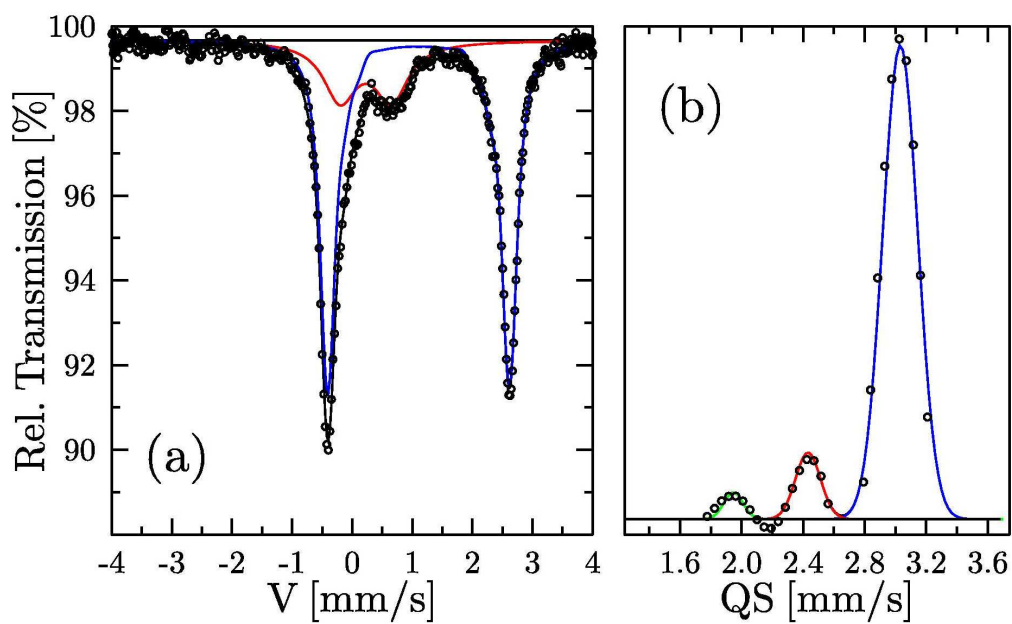

**Fig. 7**
(a) $^{57}$Fe Mössbauer spectrum recorded at 295 K on sample of Ascofer in form of powder kept for 10 months in contact with air, and (b) the corresponding distribution of the quadrupole splitting derived from the subspectrum ascribed to $Fe^{2+}$ ions (major doublet in (a)). In the latter 3 different sites are indicated

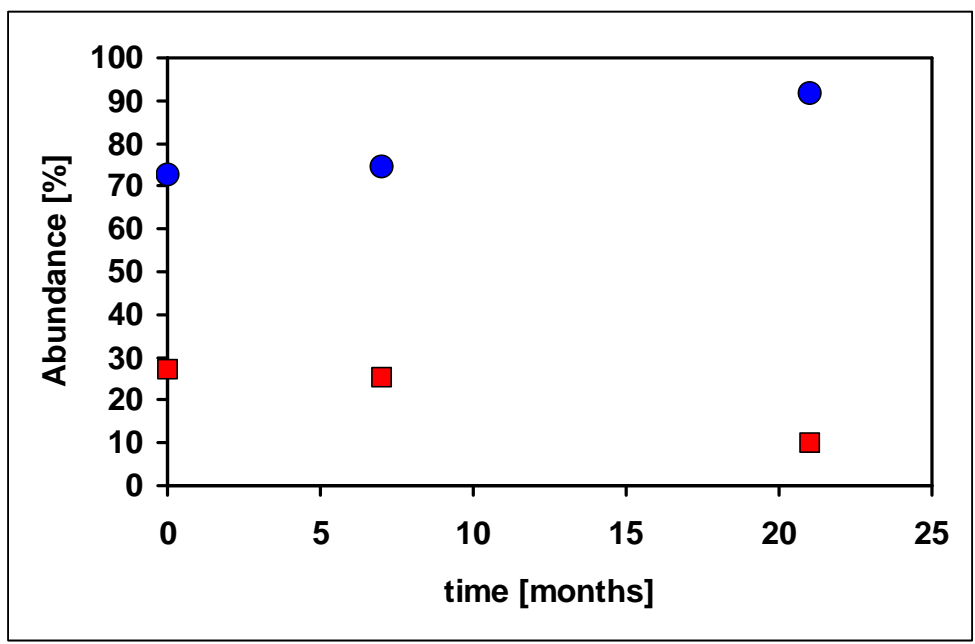

**Fig. 8**
Effect of aging on the relative abundance of $Fe^{2+}$ ions (circles) and of $Fe^{3+}$ ions (squares) in the ferrous gluconate kept in a dark and dry place.

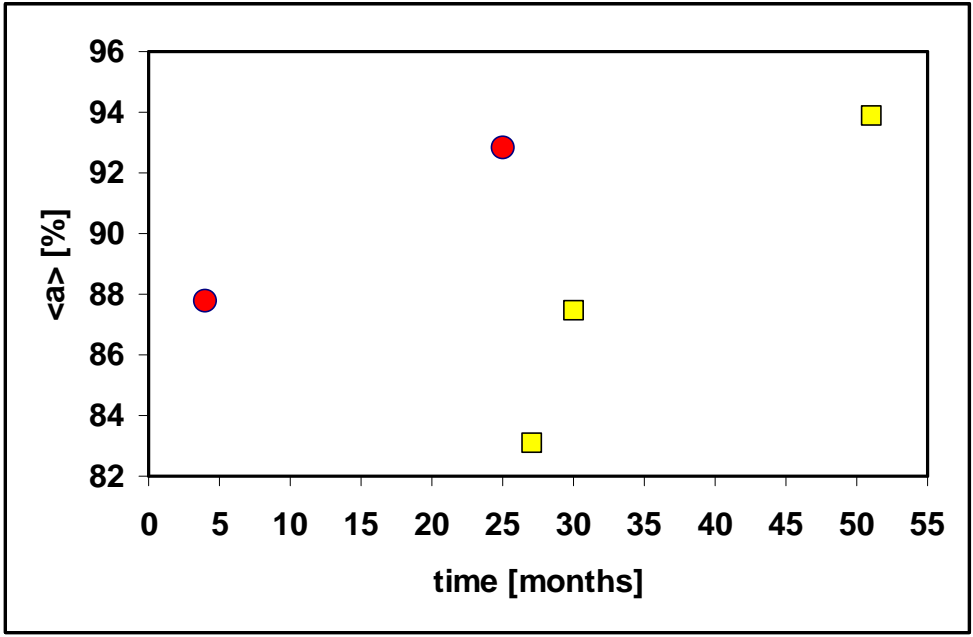

**Fig. 9**
Relative abundance of $Fe^{2+}$ ions, $<a>$, as determined in the series 150503 (squares) and in the series 150505 (circles) versus age of the samples.

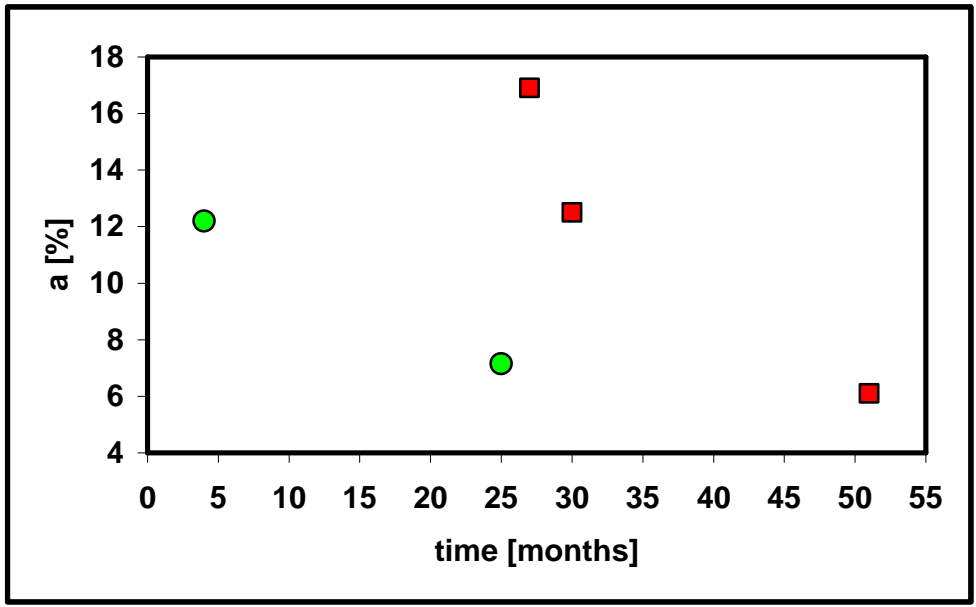

**Fig. 10**
The relative abundance of $Fe^{3+}$ ions, $a$, as determined in the series 150503 (squares) and in the series 150505 (circles) versus age of the samples.

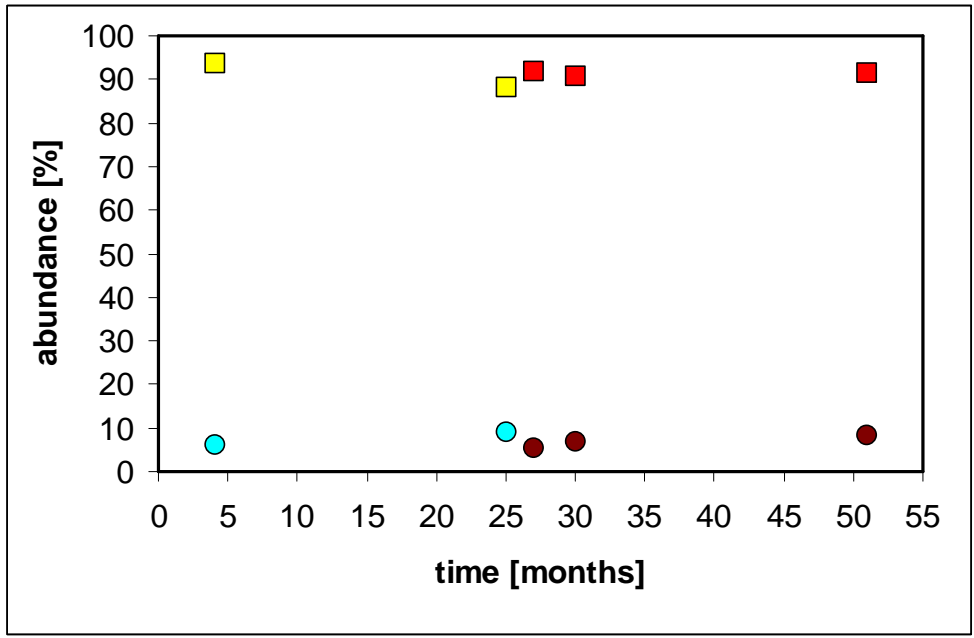

**Fig. 11**
The relative abundance of $Fe^{2+}$ ions on site 1 (squares) and site 2 (circles) versus age of Ascofer® as determined for the series 150503 and 150505.

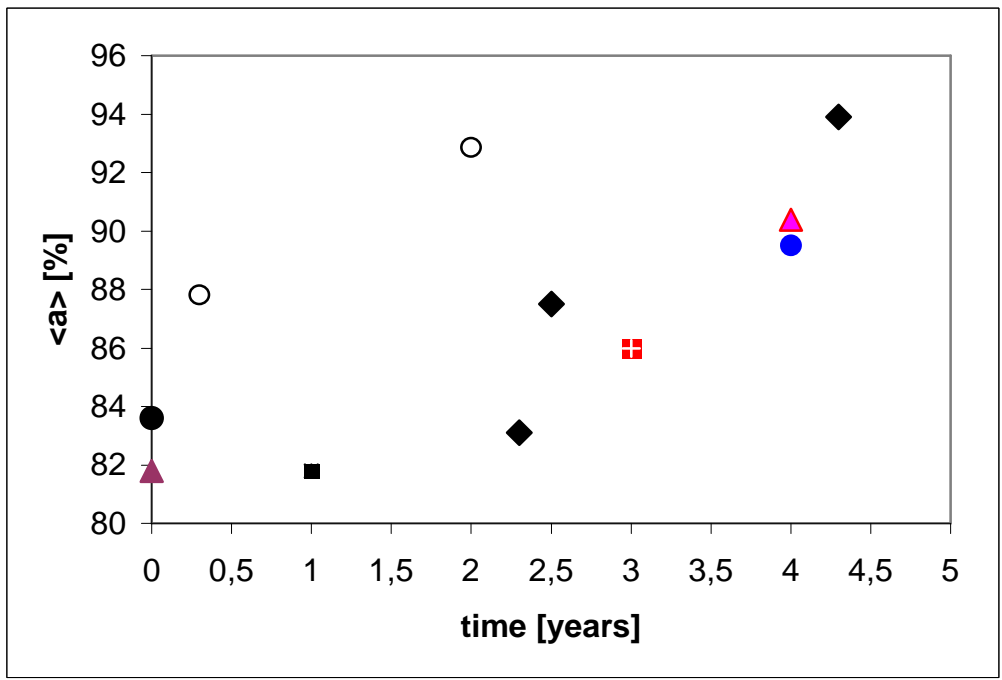

**Fig. 12**
The relative abundance of the $Fe^{2+}$ ions, $<a>$, versus age of Ascofer® samples belonging to different series. The samples from the same series are marked by the same symbol.

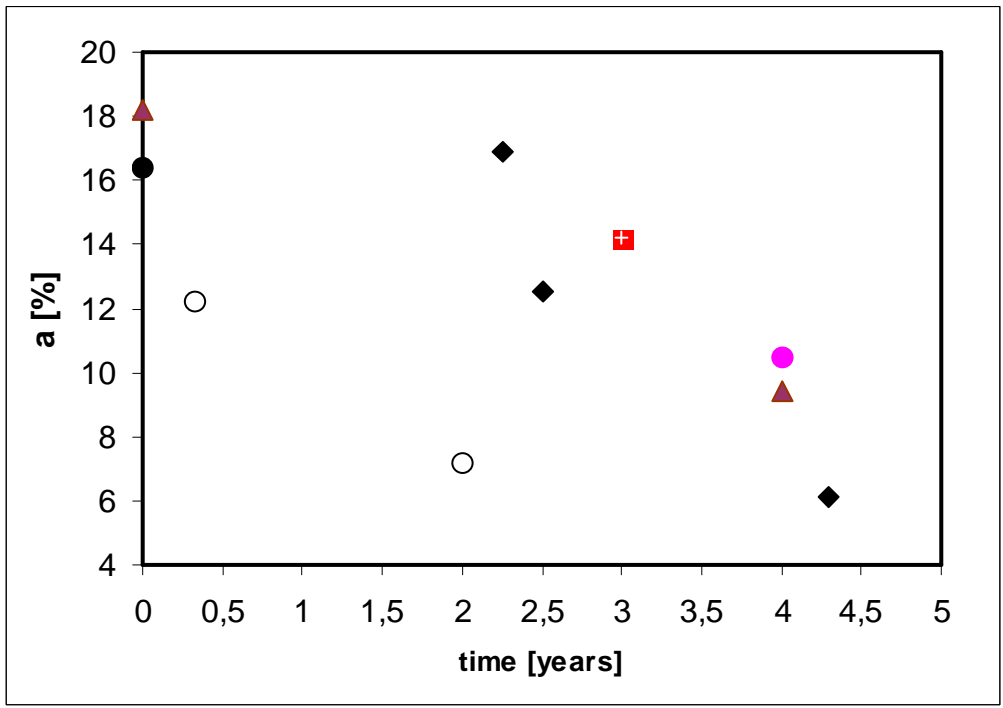

**Fig. 13**
The relative abundance of the Fe$^{3+}$ ions, *a,* versus age of Ascofer® samples belonging to different series. The samples from the same series are marked by the same symbol.